\documentclass[aps,pre,twocolumn,superscriptaddress,showpacs]{revtex4}
\usepackage{latexsym, amscd}
\usepackage{amsmath}
\usepackage{graphicx}

\begin{document}

\title{Two - dimensional solitons in media with the stripe - shaped nonlinearity
modulation}
\author{ Nguyen Viet Hung $^{1}$, Pawe\l \ Zi\'n $^{1}$, Marek Trippenbach $^{2}$, and Boris A. Malomed $^{3}$ }
\affiliation{ $^1$  Soltan Institute for Nuclear Studies, Ho\.{z}a
69,
PL-00-681 Warsaw, Poland \\
 $^{2}$ Institute of Theoretical Physics, Physics Department, Warsaw
University, Ho\.{z}a 69, PL-00-681 Warsaw, Poland\\
$^{3}$ Department of Physical Electronics, School of Electrical
Engineering, Faculty of Engineering, Tel Aviv University, Tel Aviv
69978, Israel}

\begin{abstract}
We introduce a model of media with the cubic attractive nonlinearity
concentrated along a single or double stripe in the two-dimensional (2D)
plane. The model can be realized in terms of nonlinear optics (in the
spatial and temporal domains alike) and BEC. In recent works, it was
concluded that search for stable 2D solitons in models with a spatially
localized self-attractive nonlinearity is a challenging problem. We make use
of the variational approximation (VA) and numerical methods to investigate
conditions for the existence and stability of solitons in the present
setting. The result crucially depends on the transverse shape of the stripe:
while the rectangular profile supports stable 2D solitons, its smooth
Gaussian-shaped counterpart makes all the solitons unstable. The double
stripe with the rectangular profile admit stable solitons of three distinct
types: symmetric and asymmetric ones with a single peak, and double-peak
symmetric solitons. The shape and stability of single-peak solitons of
either type are accurately predicted by the VA.
Collisions between stable solitons are briefly considered too, by means of direct simulations. 
Depending on the relative velocity we observe excitation, decay or catastrophic self focusing.
\end{abstract}

\maketitle

\section{Introduction and model}

The propensity of localized patterns (solitons) to the catastrophic self focusing in
two-dimensional (2D) media with the self-attractive nonlinearity of the most
fundamental cubic (Kerr) type is a well-known property which impedes the
creation of 2D solitons in optics and/or Bose-Einstein condensates (BECs)
\cite{1}. It was theoretically predicted for Kerr media \cite{5-8} and
demonstrated experimentally in photorefractive crystals, whose nonlinearity
is saturable \cite{2-4}, that the stabilization of 2D solitons may be
provided by periodic potentials, which can be created as optical lattices.

A different stabilization mechanism for 2D solitons was theoretically
elaborated in the form of nonlinear lattices \cite{Sivan,HS,Barcelona},
i.e., a spatially periodic or localized modulation of the nonlinearity. In
BEC this setting may be induced by dint of the Feshbach resonances
controlled through nonuniform magnetic fields \cite{SU}. In optics, similar
media may be structured as liquid-crystal-filled \cite{liquid-crystal} or
all-solid \cite{all-solid} microstructured fibers, using combinations of
materials matched in their refractive index but featuring different Kerr
coefficients. Actually, even in theoretical studies the stabilization of 2D
solitons by the spatial modulation of the cubic nonlinearity is a difficult
problem. In particular, it was not possible to find stable 2D solitons
supported by spatially periodic nonlinear lattices similar to their linear
counterpart which readily stabilize the solitons. Quasi-1D nonlinear
lattices, i.e., nonlinearity-modulation profiles which periodic functions of
a single coordinate, give rise to a tiny stability area for 2D solitons,
suggesting a conclusion that this mechanism is irrelevant to physical
applications \cite{Sivan} (while quasi-1D linear lattices easily create
stable solitons in a sizeable parameter region \cite{Q1D}). In fact, the
only nonlinear 2D structures that have thus far produced positive results as
concerns the soliton stability are represented by a circle filled with the
nonlinear material, which is embedded into a linear or defocusing nonlinear
medium \cite{HS}, and a lattice of such circles \cite{Barcelona}. In
particular, it was concluded that the nonlinearity-modulation profile with
sharp edges (such as those of the circles) provide for an essentially better
stabilization of 2D solitons than smooth profiles.

The objective of the present work is to investigate possibilities to create
\emph{stable} 2D solitons by the nonlinearity concentrated along a single or
double quasi-1D stripe, rather than patterned as a periodic lattice. It will
be demonstrated that these settings give rise to a finite stability areas
for several types of 2D solitons, on the contrary to the negligible
stabilization region supported by the quasi-1D periodic lattice \cite{Sivan}%
. The results will be obtained in a semi-analytical form using an
appropriate variational approximation (VA), and, in parallel, by means of
numerical methods.

Following Refs. \cite{HS}-\cite{Barcelona}, the general scaled form of the
2D equation with the stripe-shaped modulated nonlinearity can be written as%
\begin{equation}
i\psi _{t}=-\left( 1/2\right) \left( \psi _{xx}+\psi _{yy}\right)
+g(x)\left\vert \psi \right\vert ^{2}\psi .  \label{1}
\end{equation}%
In this work, we consider three forms of the quasi-1D modulation of
coefficient $g(x)\leq 0$ accounting for the self-attractive nonlinearity:
the Gaussian,%
\begin{equation}
g(x)=-\exp \left( -x^{2}\right) ,  \label{3a}
\end{equation}%
a single rectangular box,%
\begin{equation}
g(x)=\left\{
\begin{array}{ll}
-1 & \mbox{at }~~|x|\leq 2, \\
0 & \mbox{at }~~|x|>2,%
\end{array}%
\right.   \label{one}
\end{equation}%
and two symmetrically set boxes:

\begin{equation}
g(x)=\left\{
\begin{array}{ll}
-1 & \mbox{ at }~~d-1\leq |x|\leq d+1, \\
0 & \mbox{ at }~~|x|>d+1,~|x|<d-1.%
\end{array}%
\right.   \label{two}
\end{equation}%
In all these cases, the coefficients may be fixed to their values adopted in
Eqs.~(\ref{1})-(\ref{two}) by means of rescalings, i.e., the modulation
profiles (\ref{3a}) and (\ref{one}) have no free parameters, while their
double-box counterpart (\ref{two}) depends on a single parameter, $d>1$,
which measures half of the distance between centers of the boxes, whose
width is fixed to be $2$. Notice, that if $d<1$ we are dealing effectively with a single box.

In the application to BEC, Eq.~(\ref {1}) is the scaled Gross-Pitaevskii
equation for the mean-field wave function, $\psi \left( x,y,t\right) $ \cite%
{HS}-\cite{Barcelona}. In optics, the same model admits two different
realizations. With $t$ replaced by the propagation distance, $z$, Eq.~(\ref%
{1}) may be considered as the nonlinear Schr\"{o}dinger (NLS)\ equation to
govern the transmission of light beams in the bulk linear ambient medium,
with the transverse structure induced by an embedded nonlinear slab [Eqs.~(%
\ref{3a}) or (\ref{one})], or two parallel slabs, in the case of Eq.~(\ref%
{two}). On the other hand, the same equation (\ref{1}), with $t$ replaced by
$z$ and $y$ replaced by the temporal variable, $\tau \equiv t-z/c$, where $c$
is the group velocity of the carrier wave, describes the transmission of
spatiotemporal optical signals (alias 2D \textit{light bullets} \cite{1}) in
a planar linear waveguide, with an embedded nonlinear single or double
stripe. Accordingly, stable 2D solitons reported in this paper are
interpreted as solitary pulses of matter waves in the BEC, or spatial
solitons in the bulk optical medium, or, finally, as the light bullets
guided in the planar medium. Stationary soliton solutions are sought for as $%
\psi (x,y,t)=e^{-i\mu t}\phi (x,y)$, where $\mu $ is the chemical potential
(in terms of BEC), and function $\phi (x,y)$ satisfies equation
\begin{equation}
\mu \phi +\left( 1/2\right) \left( \phi _{xx}+\phi _{yy}\right) -g(x)|\phi|
^{2}\phi=0.  \label{2}
\end{equation}

Equation (\ref{1}) conserves the norm, $N=\int \int |\psi (x,y)|^{2}
dxdy,$ which will play the role of an intrinsic parameter of
soliton-solution families. In BEC, $N$ is proportional to the total
number of atoms, while in the bulk and planar optical waveguides it
measures, respectively, the total power or energy. Equation
(\ref{1}) also conserves
the Hamiltonian and $y$-component of the momentum,%
\begin{equation}
H= \frac{1}{2} \int \int \left[ \left( \left\vert \psi
_{x}\right\vert ^{2}+\left\vert \psi _{y}\right\vert ^{2}\right)
+g(x)|\psi |^{4}\right] dxdy,  \label{6}
\end{equation}%
\begin{equation}
P=i\int \int \left( \psi _{x}^{\ast }\psi -\psi _{x}\psi ^{\ast }\right)
dxdy.  \label{P}
\end{equation}%
For the application of the VA, it is relevant to mention that Eq.~(\ref{2})
may be derived from the Lagrangian,
\begin{equation}
L=\int \int \left[ \mu |\phi| ^{2}-\frac{1}{2}\left( |\phi
_{x}|^{2}+|\phi _{y}|^{2}\right) -\frac{1}{2} g(x)|\phi|
^{4}\right] dxdy.  \label{18}
\end{equation}

The rest of the paper is organized as follows. The solitons supported by the
Gaussian profile (\ref{3a}) are considered in Section II, where it is
demonstrated that they all are unstable. Stable solitons are found in
Sections III and IV, in the models based on the single- and double-box
modulation profiles (\ref{one}) and (\ref{two}), respectively. In these
sections, the VA and numerical methods are used in parallel. In particular,
three types of solitons are found in the setting based on the double stripe:
symmetric and asymmetric ones with a single peak, and double-peak symmetric
solitons. All the soliton species have their stability regions (for the
double-peak soliton, it is very narrow). In Section V, collisions between
solitons are studied by means of direct simulations. The paper is concluded
by Section VI.

\section{\textbf{The stripe with the Gaussian profile}}

We start the analysis by considering 2D solitons supported by the modulation
profile (\ref{3a}). First, we apply the VA based on the natural ansatz of
the Gaussian type too, with widths $a$ and $b$:
\begin{equation}
\phi (x,y)=A\exp \left[ -\left( 1/2\right) \left(
a^{2}x^{2}+b^{2}y^{2}\right) \right] ,  \label{7}
\end{equation}%
whose norm is $N=\int \int \phi ^{2}(x,y)dxdy=\pi
A^{2}/\left( ab\right) .$The substitution of the ansatz into Hamiltonian (%
\ref{6}) yields the following expression, where the squared amplitude, $A^{2}
$, is eliminated in favor of $N$:%
\begin{equation}
H=\frac{N}{4}\left( a^{2}+b^{2}-\frac{\sqrt{2}Na^{2}b}{\pi \sqrt{1+2a^{2}}}%
\right) .  \label{9}
\end{equation}%
The variational equations for parameters $a$ and $b$, $\partial H/\partial
a=\partial H/\partial b=0$ yield a system of algebraic equations for $a$ and
$b$:%
\begin{eqnarray}
\pi (1+2a^{2})^{3/2}-\sqrt{2}Nb(1+a^{2}) &=&0,  \label{10} \\
\pi b\sqrt{2\left( 1+2a^{2}\right) }-a^{2}N &=&0.  \label{11}
\end{eqnarray}%
Equation (\ref{11}) can be used to eliminate parameter $b$:
\begin{equation}
b=a^2N/\left[ \pi \sqrt{2(1+2a^{2})}\right] .  \label{b}
\end{equation}%
Substituting this into Eq.~(\ref{10}), we derive a quadratic equation for $%
a^{2}$:%
\begin{equation}
a^{2}(1+a^{2})=\left( \pi /N\right) ^{2}(1+2a^{2})^{2}.  \label{11a}
\end{equation}%
A physical (positive) root of Eq.~(\ref{11a}) is
\begin{equation}
a^{2}=\frac{N}{2\sqrt{N^{2}-4\pi ^{2}}}-\frac{1}{2},  \label{11b}
\end{equation}%
which predicts that the 2D solitons exist provided that the norm exceeds a
threshold value, $N_{\mathrm{thr}}=2\pi $. In fact, the same generic feature
of solitons was found in other models based on the spatial modulation of the
nonlinearity, both two- \cite{HS,Sivan,Barcelona} and one-dimensional \cite%
{1D}.

Further, we insert $b$ and $a$ from Eqs.~(\ref{b}) and (\ref{11b}) into Eq.~(%
\ref{9}) to obtain the expression for the Hamiltonian predicted by the VA:%
\begin{equation}
H=\frac{N}{16\pi ^{2}}\left[ N\left( N-\sqrt{N^{2}-4\pi ^{2}}\right) -2\pi
^{2}\right] .  \label{12}
\end{equation}%
Using the definition of the chemical potential, $\mu =dH/dN$, the
Vakhitov-Kolokolov (VK) criterion \cite{VK}, $d\mu /dN<0$, which is assumed
to be a necessary condition for the stability of the solitons, is then cast
in the form of
\begin{equation}
d^{2}H/dN^{2}<0.  \label{14}
\end{equation}%
From (\ref{12}) we obtain
\begin{equation}
\frac{d^{2}H}{dN^{2}}=\frac{1}{8\pi ^{2}}\left[ 3N+\frac{16\pi
^{2}N^{2}-3N^{4}-16\pi ^{4}}{(N^{2}-4\pi ^{2})^{3/2}}\right] .  \label{15}
\end{equation}%
It is easy to check that this expression is always positive, hence the VA
predicts that the 2D solitons supported by the stripe-shaped modulation of
the attractive nonlinearity with the Gaussian profile \emph{cannot be stable}. We also calculated the matrix of second order derivatives of the Hamiltonian with respect to $a$ and $b$ and check that the stationary point (solution of Eqs.~(\ref{10}) and (\ref{11})) is not a minimum. Direct simulations (not shown here) corroborate this prediction: no stable
2D solitons can be found in the model based on Eqs.~(\ref{1}) and (\ref{3a}).

\section{\textbf{The modulation stripe with the box-shaped profile}}

\label{onesquare}

\subsection{The variational approximation}

Proceeding to the model with the rectangular modulation profile (\ref{one}),
we again start with the VA, using two different \textit{ans\"{a}tze},
namely, the same Gaussian-shaped one as in Eq.~(\ref{7}), and also the
product of hyperbolic secants:
\begin{equation}
\phi (x,y)=A~\mathrm{sech}(x/a)\mathrm{sech}(y/b),  \label{17}
\end{equation}%
whose norm is $N=4A^{2}ab$. An ansatz of the latter type was
originally used for modeling multidimensional solitons in Ref.
\cite{Japan}.

\begin{figure}[h]
\includegraphics[width=9cm]{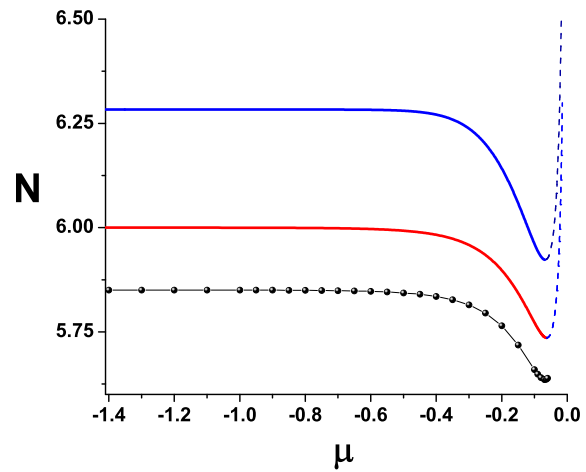}
\caption{(Color online) Curves $\protect\mu (N)$ for the soliton family, as
predicted by the variational approximations based on the sech-sech ansatz (%
\protect\ref{17}) (red) and Gaussian (\protect\ref{7}) (bue), in the
\ model with the single-box modulation profile (\protect\ref{one}).
The VK-stable portions of the family are shown by solid curves.
Numerically found stable solitons are shown by the chain of
circles.} \label{hin1h}
\end{figure}

The substitution of ansatz (\ref{17}) into Lagrangian (\ref{18}) yields
\begin{eqnarray}
L&=&4A^{2}ab\mu -\frac{2A^{2}(a^{2}+b^{2})}{3ab}\nonumber\\
&+&\frac{2}{9}abA^{4}\mathrm{sech}^{3}\left( \frac{2}{a}\right) %
\left[ 3{\sinh }\left( \frac{2}{a}\right) +{\sinh }\left(
\frac{6}{a}\right) \right].\nonumber\\
\label{19}
\end{eqnarray}%
Deriving the Euler-Lagrange equations from this expression, it is
possible to eliminate $A$ and $b$, ending up with equations
\begin{equation}
\mu =\frac{1}{3a^{2}}\left[ \frac{36}{24+4a~{\sinh }(\frac{4}{a})+a~{%
\sinh }(\frac{8}{a})}-1\right] ,  \label{20}
\end{equation}%
\begin{equation}
N=\frac{12a~{\cosh }^{2}(\frac{2}{a}){\coth }(\frac{2}{a})}{\left[ 2+%
{\cosh }(\frac{4}{a})\right] \sqrt{a\left\{ a+24\left[ 4{\sinh }
(\frac{4}{a})+{\sinh }(\frac{8}{a})\right] ^{-1}\right\} }}.  \label{21}
\end{equation}%
The system of Eqs.~(\ref{20}) and (\ref{21}) was solved numerically. Curves $%
N(\mu )$ for the soliton families, produced by the \textit{ans\"{a}tze} of
both types, are shown in Fig. \ref{hin1h}. According to the VK criterion,
portions of the soliton families depicted by the solid curves may be \emph{%
stable}.

The Hamiltonian corresponding to ansatz (\ref{17}) is
\begin{equation}
H=\frac{N}{6}\left( \frac{1}{a^{2}}+\frac{1}{b^{2}}\right) -\frac{N^{2}}{36ab%
}\left[ 2+\mathrm{sech}^{2}\left( \frac{2}{a}\right) \right] {\tanh }%
\left( \frac{2}{a}\right) ,  \label{19aa}
\end{equation}%
cf. Eq.~(\ref{12}). Again the stability condition obtained from Fig.~\ref{hin1h} using VK criterion coincide with the fact that the  Hamiltonian (\ref{19aa}) has a local minimum.

\subsection{Numerical results}

We have performed systematic numerical simulations of Eqs.~(\ref{1}), (\ref%
{one}), with the aim to verify the predictions of the VA for this model.
First, stationary shapes of 2D solitons were generated by means of
imaginary-time simulations. Then, their stability was checked using the
split-step Fourier-transform method for simulating the evolution of
perturbed solitons, to which a small disturbance was added. The initial
perturbation changed the total norm of the soliton by up to $0.1\%$ (adding
perturbations at this level were sufficient to clearly distinguish between
stable and unstable solitons).

The simulations produce \emph{stable} 2D solitons in the interval of $%
5.645\leq N\leq 5.85$. The comparison of the numerical findings with
the predictions of the VA is shown in the Figure \ref{hin1h}.
\\
The fact that the replacement of the smooth Gaussian modulation
profile by the box-shaped one agrees with the above-mentioned
general trend discovered
in other models dealing with the spatially modulated nonlinearity, \textit{%
viz}., that modulation profiles with \emph{sharp edges} are more apt to
generate stable 2D solitons than their smoothly shaped counterparts \cite%
{HS,Barcelona}.

\section{\textbf{The double-box modulation profile}}

\subsection{The variational approximation}

To consider the model based on the dual-box modulation profile as given by
Eq.~(\ref{two}), we have started with the VA based on the factorized-sech
ansatz similar to that introduced above in the form of Eq.~(\ref{17}), but
with an additional degree of freedom, $c$, which makes it possible to
consider solitons configurations with a spontaneously broken symmetry:

\begin{equation}
\phi (x,y)=\frac{1}{2}\sqrt{\frac{N}{ab}}\mathrm{sech}\left( \frac{x-c}{a}%
\right) \mathrm{sech}\left( \frac{y}{b}\right) .  \label{22}
\end{equation}%

\begin{figure}[h]
\includegraphics[width=4cm]{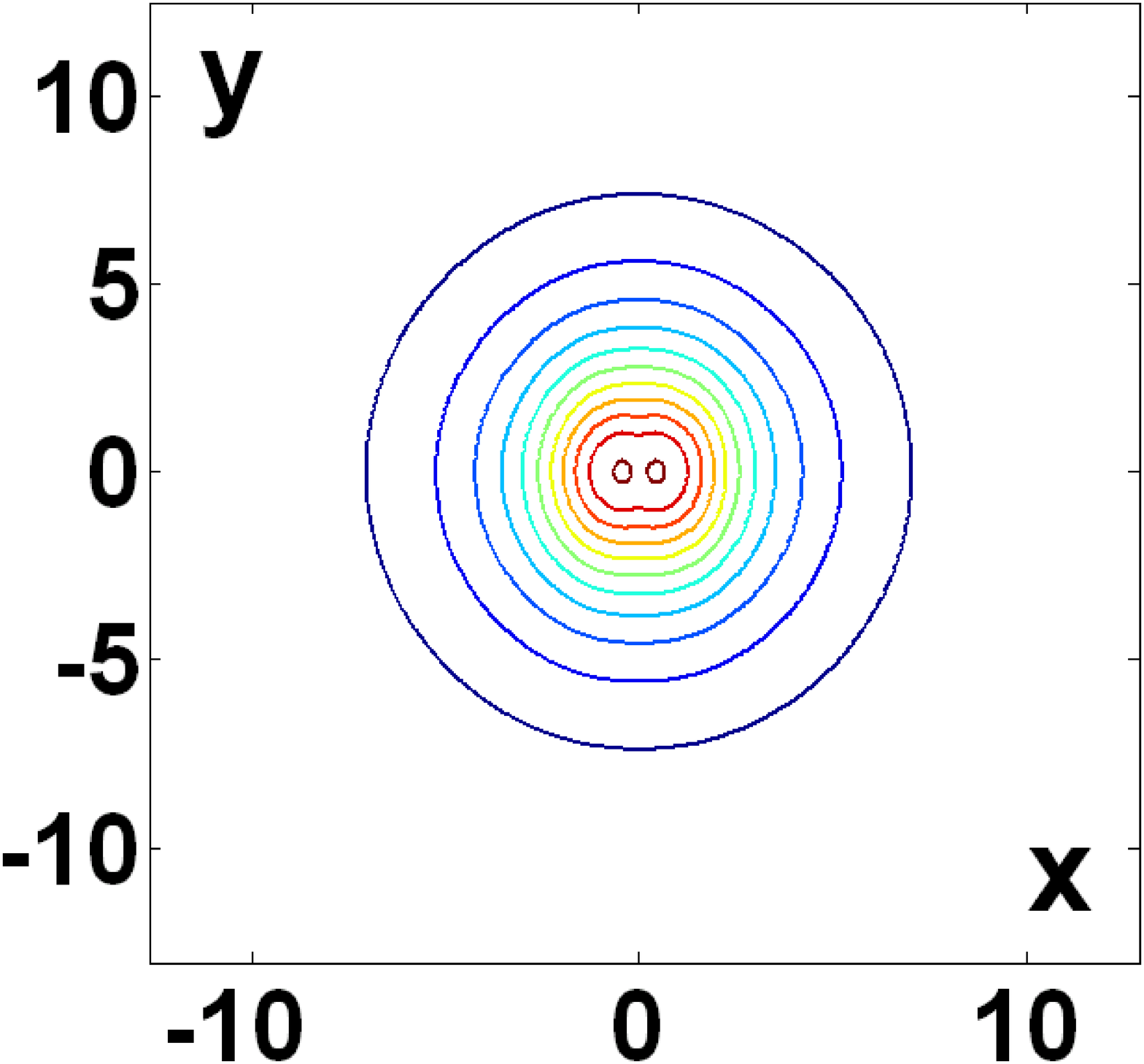}
\includegraphics[width=4cm]{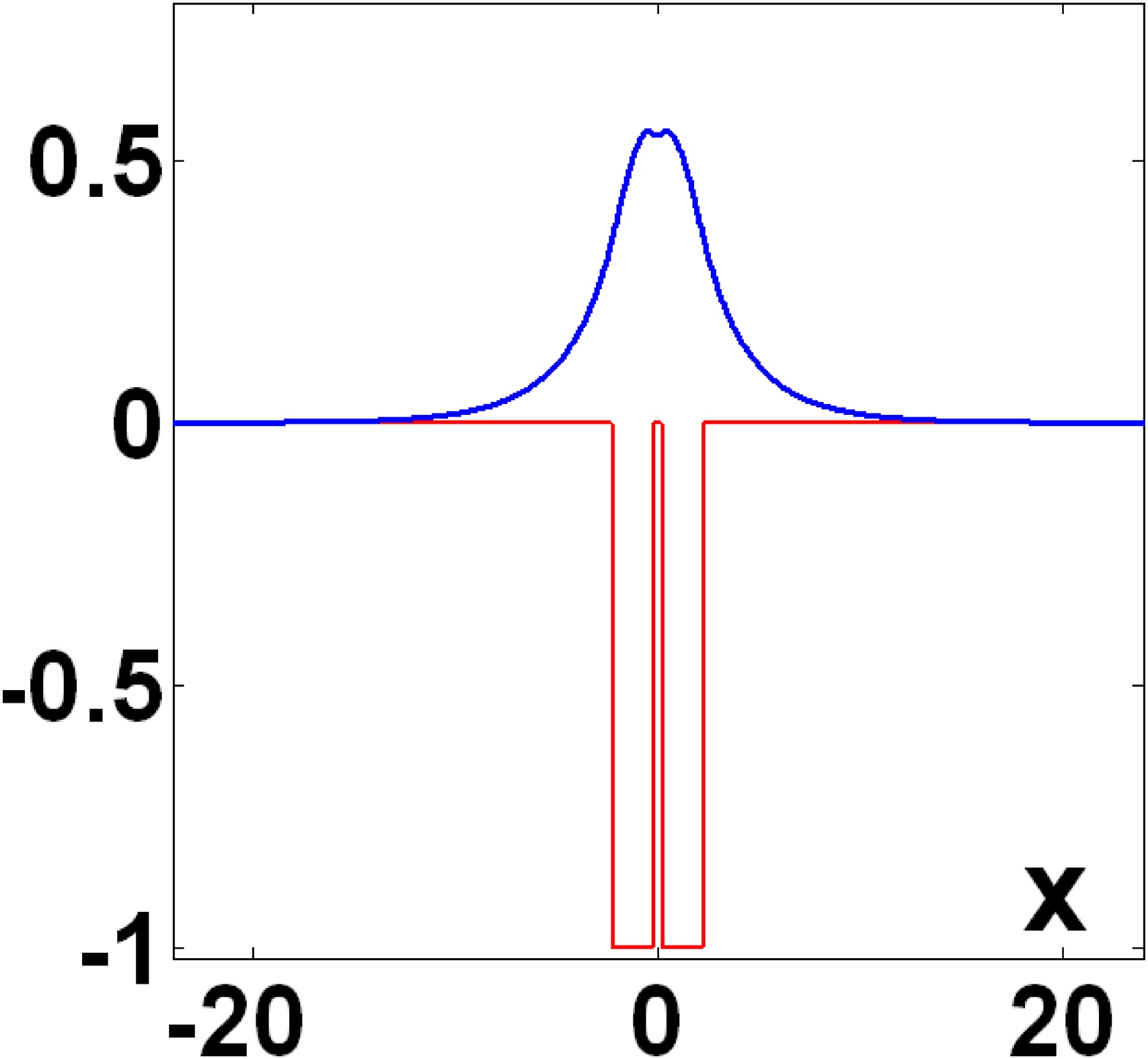}
\caption{(Color online) An example of a stable double-peak symmetric
soliton, with $N=6.82$, $\protect\mu =-0.05$, and $d=1.2$. Left: A top view
shown by means of contour plots; Right: the section drawn through $y=0$.}
\label{hin2haha}
\end{figure}

The respective expression for Hamiltonian (\ref{6}) is:%
\begin{widetext}
\begin{eqnarray} \nonumber
H &=& \frac{N^2}{72ab} \left[  \left( 2+ \mathrm{sech}^{2}\left(  \frac{d+1-c}{a} \right) \right) {\tanh }\left( \frac{c-1-d}{a}\right) -\left( 2+\mathrm{sech}^{2}\left( \frac{c+1-d}{a}\right) \right) {\tanh }\left( \frac{c+1-d}{a}\right) \right.
\\ \nonumber
&+& \left. \left( 2+\mathrm{sech}^{2}\left( \frac{d+c-1}{a}\right) \right) {\tanh}\left( \frac{c+d-1}{a}\right)
 -\left( 2+\mathrm{sech}^{2}\left( \frac{c+1+d}{a}\right) \right) {\tanh }\left( \frac{c+1+d}{a}\right)
\right] +\frac{N}{6}\left( \frac{1}{a^{2}}+\frac{1}{b^{2}}\right) .\\
\label{complex}
\end{eqnarray}
\end{widetext}
In the model with the dual-trough linear potential, formally similar to its
nonlinear counterpart (\ref{two}), the spontaneous symmetry breaking of 2D
solitons was studied in detail in Ref. \cite{we}. In a 1D model with the
nonlinearity modulation represented by two symmetric delta-functions, or by
a superposition of two Gaussians, symmetry-breaking localized states were
investigated in Ref. \cite{Dong}.

The variational equations following from expression (\ref{complex}) are
cumbersome, therefore they are not explicitly written here. These equations
were solved numerically, for both symmetric ($c=0$) and asymmetric ($c\neq 0$%
) configurations. The stability condition following from the VA was
implemented as a condition that the Hamiltonian must attain a local minimum
at stationary points. The results are presented below, in comparison with
respective numerical findings obtained from Eqs.~(\ref{1}) and (\ref{two}).

\subsection{Basic types of two-dimensional solitons. Numerical results.}

Numerical solutions for stationary 2D solitons were obtained by means of two
numerical methods, namely, the imaginary-time integration (as above), and
the spectral renormalization method \cite{MZ}. In the former case, we fixed $%
N$ and aimed to find the corresponding value of $\mu $, while the latter
algorithm was used to find $N$ corresponding to given chemical potential $%
\mu $.

Soliton solutions of three types have been found in this way: (i) symmetric
modes with a single peak; (ii) asymmetric solutions with a single peak; and
(iii) symmetric solitons with a \textit{double peak}. Solutions with two
asymmetric peaks have not been found. Obviously, modes of type (iii) cannot
be predicted by the variational ansatz (\ref{22}) (a modification of the
ansatz that would incorporate double-peak patterns leads to an extremely
messy algebra).

\subsubsection{Symmetric solitons with a single peak}

In the case of $d\leq 1$, the two boxes in modulation profile (\ref{two})
actually merge into one, bringing us back to profile (\ref{one}) that was
considered above. Accordingly, symmetric single-peak solitons are found in
this case. At $d>1$, there is a small region where the VA predicts stable
solutions. However, the full numerical solutions do not reveal any
stationary modes of this type.

\subsubsection{Double-peak symmetric solitons}

\begin{figure}[h]
\includegraphics[width=9cm]{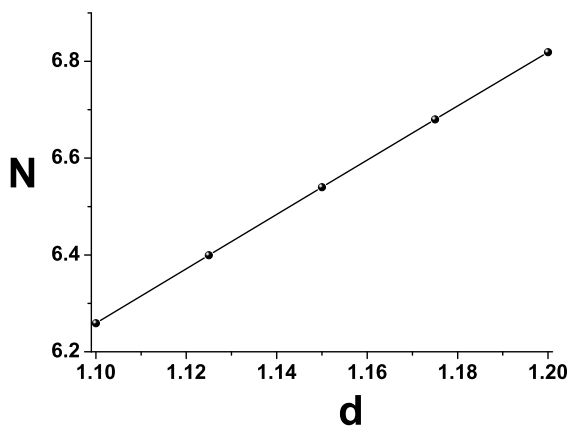}
\caption{(Color online) The norm of the stable symmetric double-peak
solitons versus the half-distance, $d$, between the two boxes in modulation
profile (\protect\ref{two}), for a fixed chemical potential, $\protect\mu %
=-0.05$.}
\label{hin3hg}
\end{figure}

In the interval of $1<d\leq 1.2$, the spectral renormalization method \cite%
{MZ} has produced stationary symmetric solutions with double peaks and a
very shallow local minimum between them, see Fig.~\ref{hin2haha} (the
imaginary-time method does not converge in this region). These solitary
modes are stable only at very small values of $|\mu |$, when the solitons
are very broad. For example, fixing $\mu =-0.05$, we have found that \emph{%
stable} double-peak symmetric solutions exist in the interval of $%
1.1\leq d\leq 1.2$, where the dependence of $N$ on $\left( d-1\right) $
turns out to be linear, see Fig.~\ref{hin3hg}. It
may be relevant to note that, because the stability of the solitons was
identified by means of direct simulations, it may happen that the stable
solitons are, strictly speaking, subject to an extremely weak instability.
Nevertheless, they are definitely stable modes in terms of physical
applications.


\subsubsection{Asymmetric single-peak solutions}

\begin{figure}[h]
\includegraphics[width=4cm]{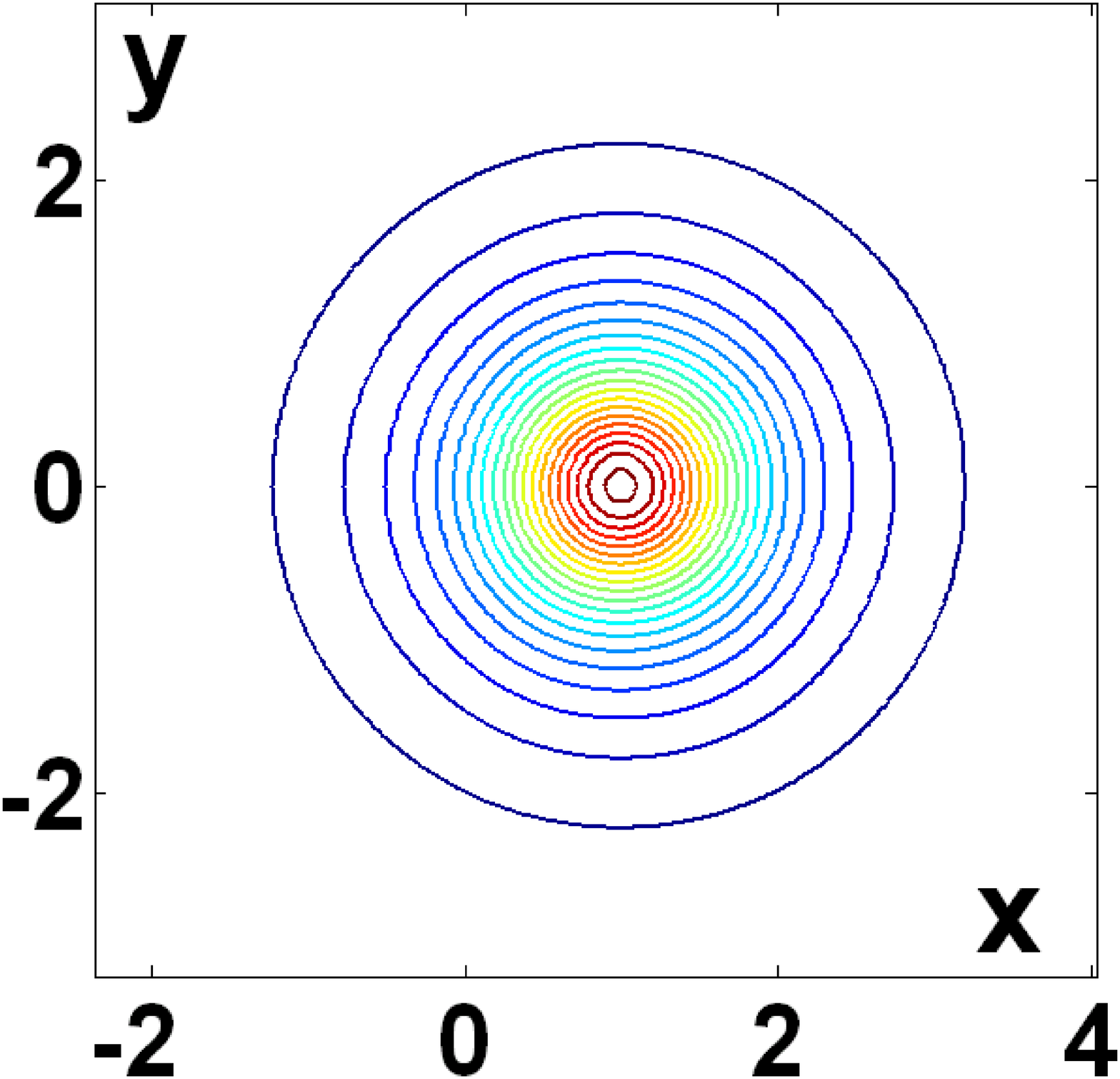}
\includegraphics[width=4cm]{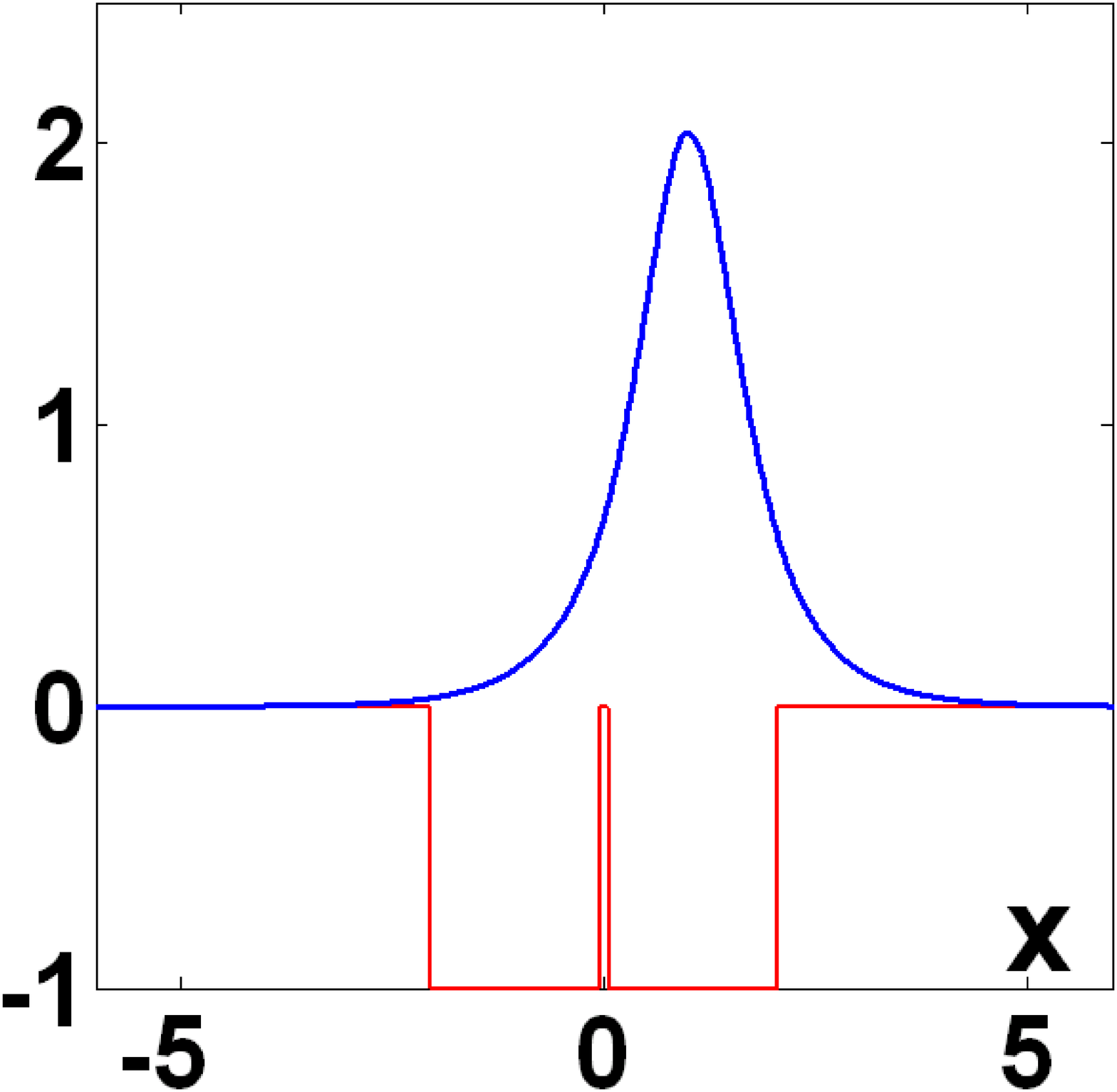}
\caption{(Color online) The same as in Fig.~\protect\ref{hin2haha}, but for
a stable asymmetric soliton with $N=5.8$, $\protect\mu =-0.84$, and $d=1.05$%
. The substrate (red) line in the bottom panel represents the two-box
profile (\protect\ref{two}).}
\label{hin3haha2}
\end{figure}

\begin{figure}[h]
\includegraphics[width=4cm]{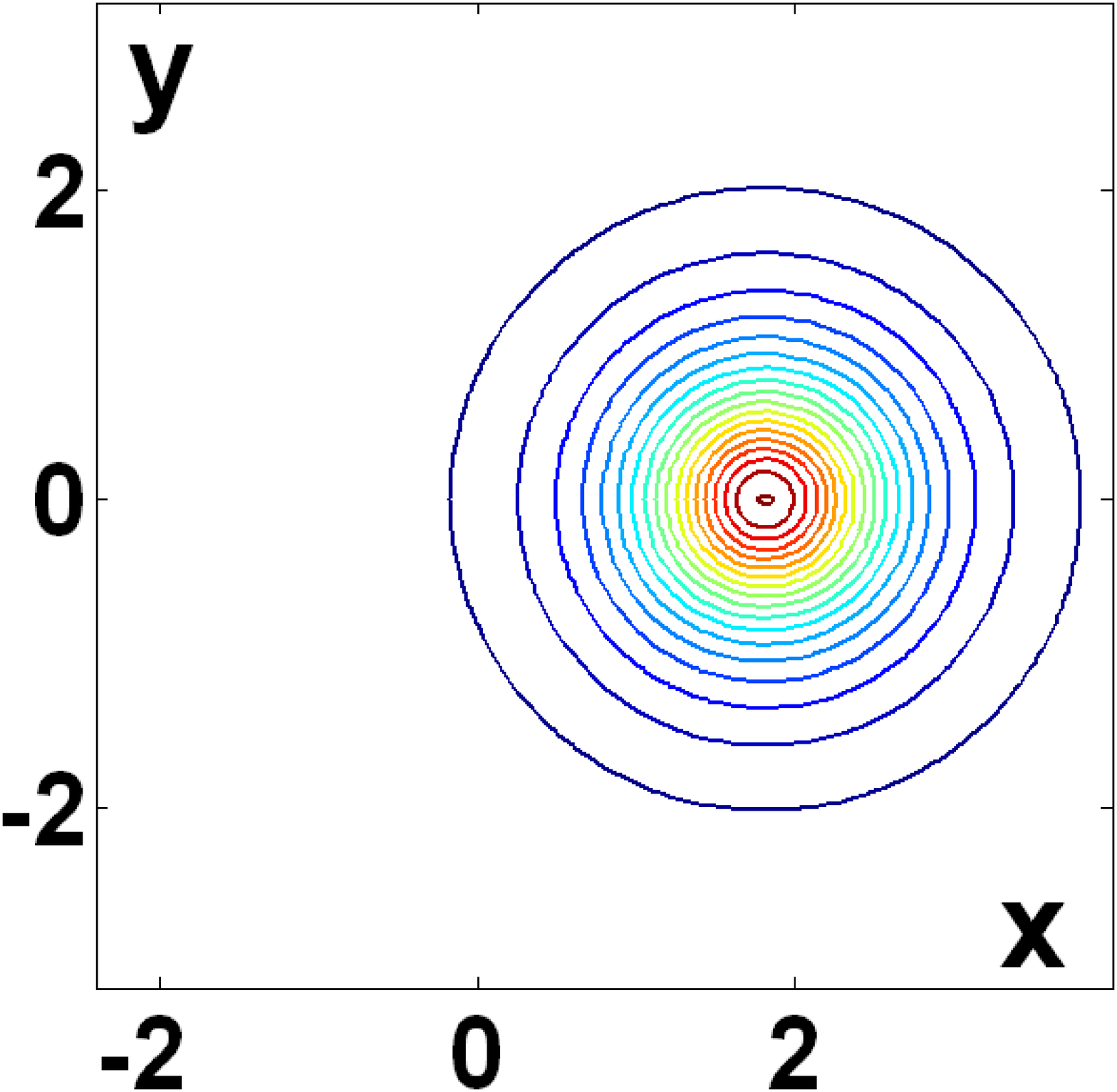}
\includegraphics[width=4cm]{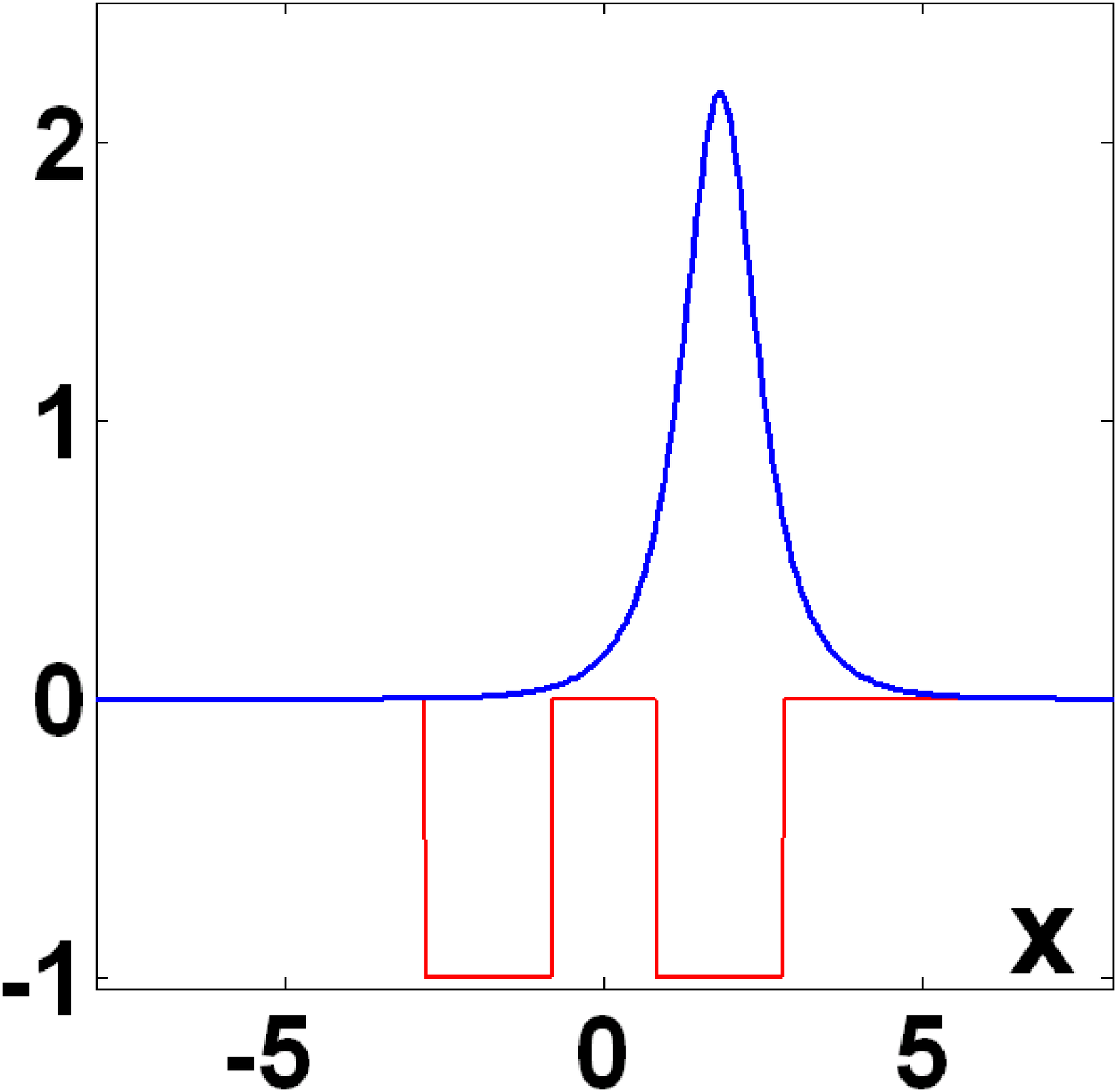}
\caption{The same as in Fig.~\protect\ref{hin3haha2}, but for a
stable asymmetric soliton with $N=5.8$, $\protect\mu =-0.962$, and
$d=1.8$.} \label{hin3haha}
\end{figure}

The system with the double-box modulation profile, given by Eq.~(\ref{two})
with $d>1$, can support asymmetric solutions with a single peak. Figures %
\ref{hin3haha2} and \ref{hin3haha} display the form of stationary wave
functions in two cases, with $d$ close to and far from $1$, respectively.

\begin{figure}[h]
\includegraphics[width=8cm]{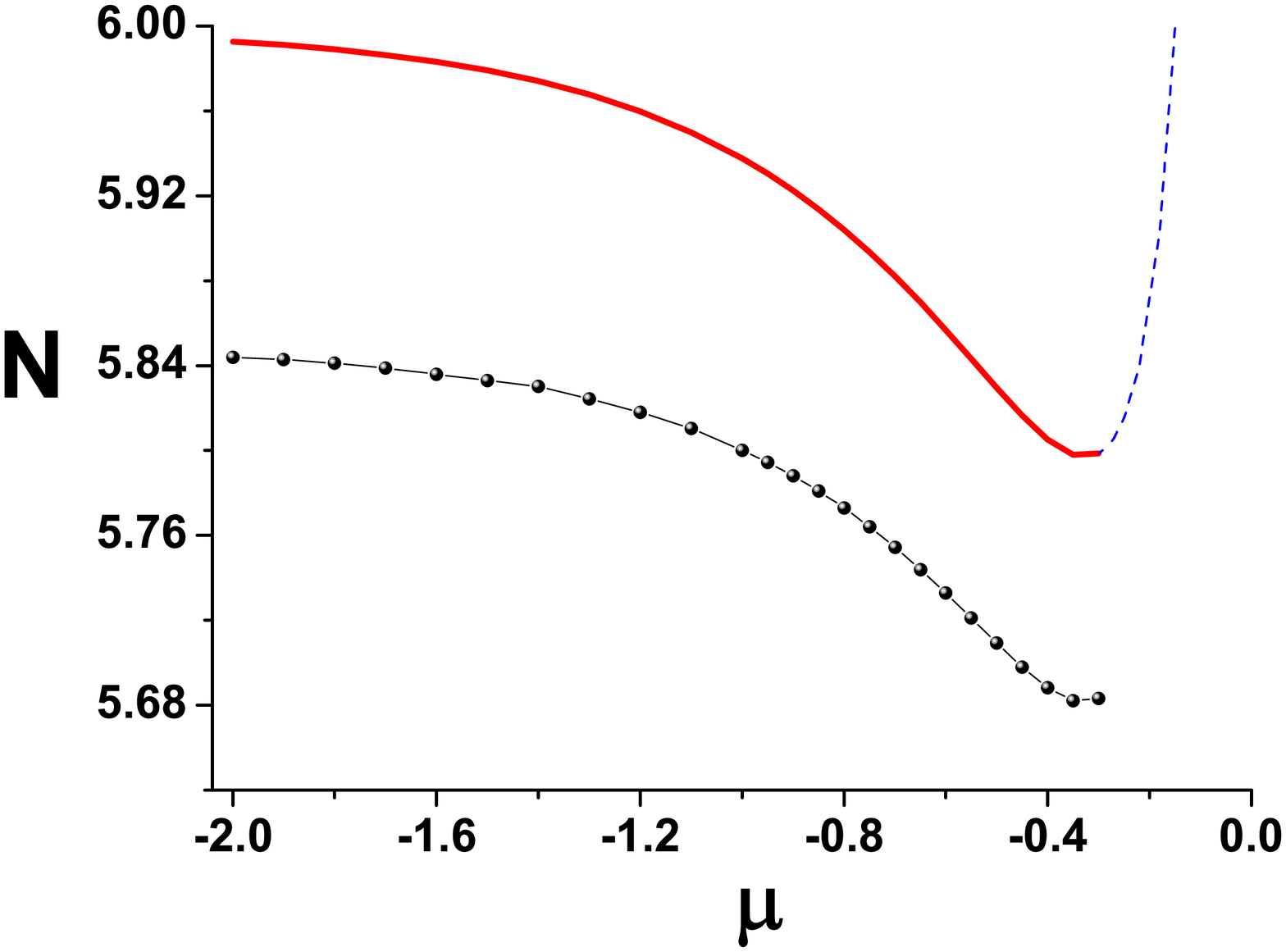}
\includegraphics[width=8cm]{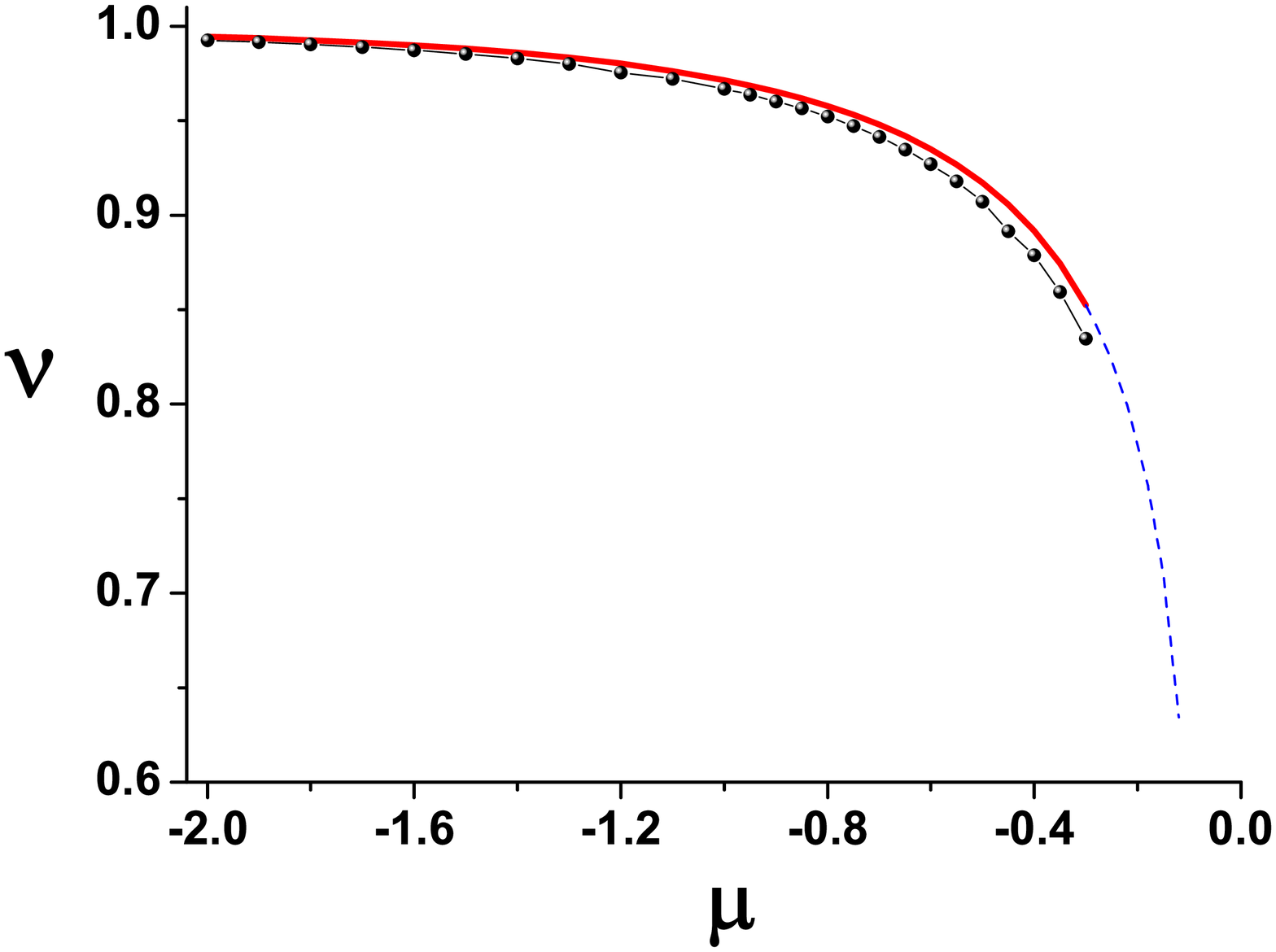}
\caption{(Color online) Curves showing the norm and asymmetry parameter, $N(%
\protect\mu )$ and $\protect\nu (\protect\mu )$, for asymmetric single-peak
2D solitons in the case of $d=1.2$. We display the comparison between the
results predicted by the variational approximation (continuous/dashed
curves, which depict stable/unstable portions of the soliton family), and
their numerical counterparts (chains of circles).}
\label{hin3h12}
\end{figure}

The asymmetry of solutions may be naturally characterized by parameter
\begin{equation}
\nu =\frac{\int_{-\infty }^{\infty }{d}y\int_{-\infty }^{0}%
{d}x|\phi (x,y)|^{2}-\int_{-\infty }^{\infty }{d}%
y\int_{0}^{\infty }{d}x|\phi (x,y)|^{2}}{N}.  \label{24}
\end{equation}%
In Fig.~\ref{hin3h12} we present curves $N(\mu )$ and $\nu (\mu )$ for the
asymmetric solitons in the case of $d=1.2$. From these plots, and similar
ones obtained at other values of $d$, it is concluded that the VA based on
ansatz (\ref{22}) provides for a good accuracy, in the comparison with
numerical results.

\subsection{Stability and symmetry-breaking diagrams}

\begin{figure}[h]
\begin{center}
\includegraphics[width=8cm]{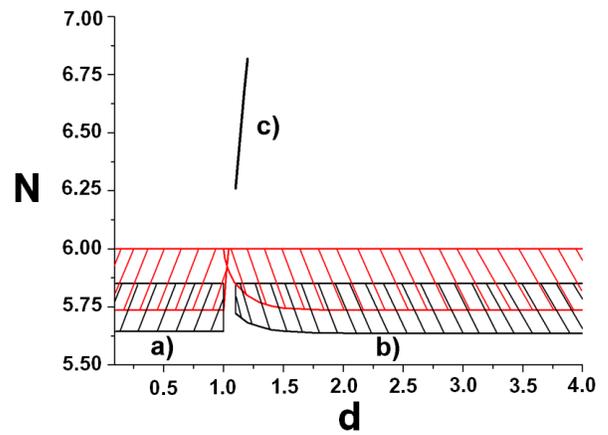}
\end{center} \caption{(Color online) The full stability diagrams for
2D solitons in the plane of $(d,N)$. In areas a) and b), stable
solutions are, respectively, symmetric and asymmetric single-peak
solitons (black and red areas depict the stability areas as
produced, severally, by the numerical and variational methods). In a
very thin area c), stable symmetric double-peak solutions have been
found, in the numerical form.} \label{hin4hg}
\end{figure}

\begin{figure}[h]
\begin{center}
\includegraphics[width=7.5cm]{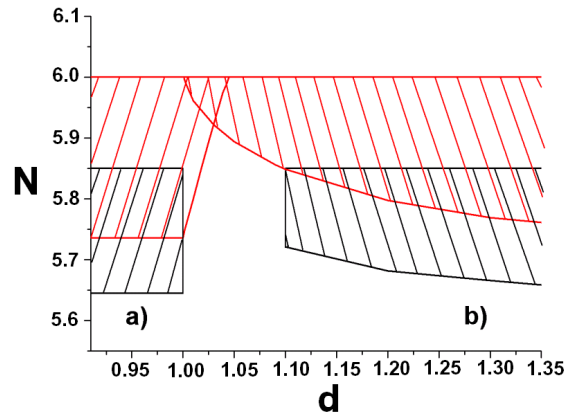}
\end{center} \caption{(Color online) A zoom of the stability diagram
from Fig.~\protect
\ref{hin4hg} close to $d=1$, where the single-box modulation profile (%
\protect\ref{one}) goes over into the double-box shape (\protect\ref{two}).}
\label{hin4hgi}
\end{figure}

The results produced by the VA and gleaned from full numerical solutions are
collected in the form of stability diagrams displayed in Figs.~\ref{hin4hg}
and \ref{hin4hgi}. It is concluded that the numerically found stability
areas (covered by black oblique lines in the figures) are shifted somewhat
downward, against the variational predictions (covered by red oblique
lines). Around $d=1$, there is a gap in the numerically found stability
areas for the symmetric and asymmetric single-peak solitons. It is extremely
difficult to fill this gap using direct numerical solutions, as the
simulations necessary to check both the existence and stability of the
solutions turn out to be very long in this case.

\begin{figure}[h]
\includegraphics[width=9cm]{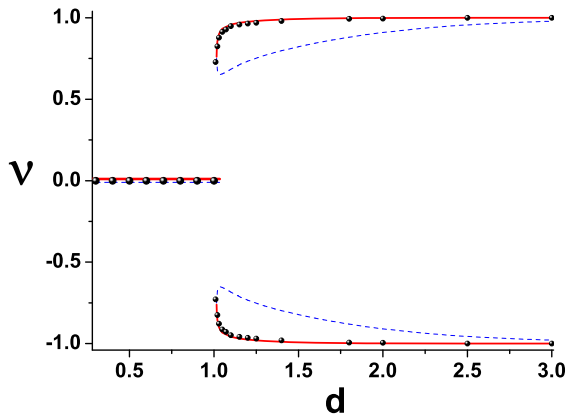}
\caption{(Color online) The symmetry-breaking diagram, displayed as the $%
\protect\nu (d)$ curve for a fixed norm of symmetric and asymmetric
single-peak solitons, see the text. The red (continuous) and blue (dashed)
curves depict, respectively, stable and unstable solutions predicted by the
VA, while chains of circles represent numerically found stable solutions.}
\label{hin4h}
\end{figure}

\begin{figure}[tbp]
\includegraphics[width=9cm]{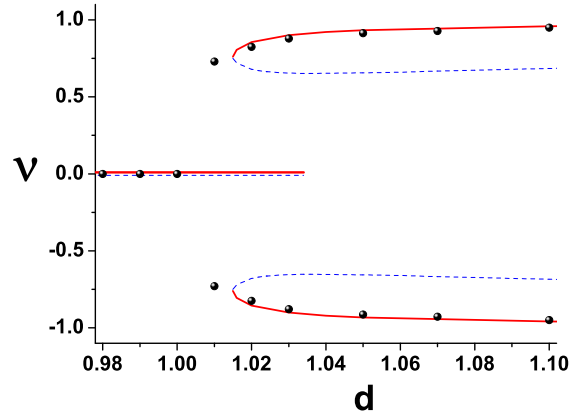}
\caption{(Color online) A blow-up of the symmetry-breaking region from Fig. %
\protect\ref{hin4h}.}
\label{hin4hiK}
\end{figure}

Further, a \textit{symmetry-breaking} diagram is displayed in Figs.~\ref%
{hin4h} and \ref{hin4hiK}. To produce this plot, we fixed the norm of the
solitons and varied the half-distance, $d$, between the two wells. This
implies that we keep a constant strength of the nonlinearity, while making
the coupling between the two box-shaped stripes weaker. The norm of the
solutions corresponding to the chain of circles in Figs.~\ref{hin4h} and %
\ref{hin4hiK}, which were found as numerical solutions to Eqs.~(\ref{1}), (%
\ref{two}), was set as $N_{\mathrm{num}}=5.8$. The constant norm of the
variational solutions presented in the same figures was fixed to be slightly
different, $N_{\mathrm{VA}}=5.95$, as Fig.~\ref{hin1h}
suggests a correction ratio, $6/5.85\approx 5.95/5.8$ between the
VA-predicted solutions and their numerically found counterparts.

The detailed picture of the symmetry-breaking transition, displayed in 
Fig.~\ref{hin4hiK} demonstrates that the VA predicts the transition of a
slightly \textit{subcritical} type, with a narrow bistability region
(coexistence of stable symmetric and asymmetric states) observed at $%
1.015<d<1.034$ (cf. Ref.~\cite{we}, where the subcritical transition
and bistability were demonstrated in the 2D model with the
double-trough linear potential). However, after reaching the
destabilization point at $d=1.034$, the line of the symmetric
solitons \emph{does not} continue to larger values of $d$, as one
might ``naively" expect. Instead, the destabilized branch of the
symmetric solitons turns back, running parallel to the stable one.
The latter feature corresponds to the existence of the pair of
stable and unstable solution branches in the model with the
single-box modulation profile, see Fig.~\ref{hin1h}. These features
of the
symmetry-breaking diagram somewhat resembles those reported in Ref.~\cite%
{Zeev}, which was dealing with a 2D model combining the double-trough linear
potential and the cubic-quintic nonlinearity. As concerns the stable and
unstable VA-predicted branches of the asymmetric solutions, it may be
expected that they will meet in the limit of large $d$, as in that limit
each of the two boxes becomes equivalent to the one in the model with the
single-box modulation profile.

Numerical solutions for symmetric solitons have also been found at some
values of $d>1$. However, the corresponding points are not shown in Fig.~%
\ref{hin4hiK}, as it was too difficult to resolve their stability.

\begin{figure}[tbh]
\includegraphics[width=9cm]{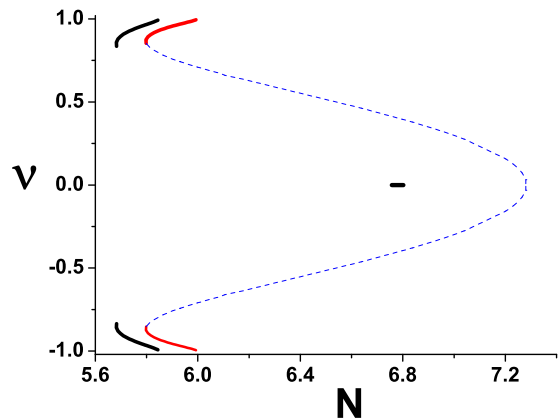}
\caption{(Color online) The symmetry-breaking and stability diagram
represented by the $\protect\nu (N)$ curve for fixed $d=1.2$. The
red (solid) and blue (dashed) curves depict, respectively, stable
and unstable subfamilies of asymmetric single-peak solitons, as
generated by the VA. The segments of black curves at the top and
bottom represent stable solitons of the same type, obtained in the
numerical form. The short black segment at the center depicts a
family of stable symmetric double-peak solutions.} \label{hin4gh}
\end{figure}

\begin{figure}[tbh]
\includegraphics[width=8cm]{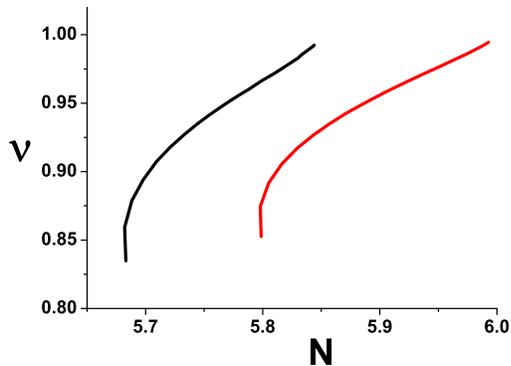}
\caption{(Color online). A zoom of the part of Fig.~\protect\ref{hin4gh} which shows in detail the variational (the
red curve) and numerical (the black curve) families of stable
asymmetric single-peak solitons.} \label{hin4g}
\end{figure}

Another representative stability and symmetry-breaking diagram can be
constructed by varying the norm, $N$ (i.e., the nonlinearity strength, as a
matter of fact) at a fixed value of the half-distance $d$ between the boxes.
This diagram and a zoom of its most essential area are displayed in Figs.~%
\ref{hin4gh} and \ref{hin4g}. The narrow interval of values of $N$ in
which stable solitons, supported by the spatially modulated nonlinearity,
may be found, is a generic feature of 2D models \cite{HS,Sivan,Barcelona}. A
noteworthy feature of this diagram is the absence of symmetric single-peak
solutions (while their stable double-peak counterparts exist in a very
narrow interval). The stable asymmetric solitons are obviously generated by
saddle-node (subcritical) bifurcations. Unlike the 1D model with the
double-stripe nonlinearity modulation \cite{Dong}, here the existence
regions for the stable solitons is bounded from above, in terms of $N$, by
the catastrophic self focusing.

\section{Collisions between moving solitons}

In the last paragraph we report our studies on the soliton
collisions. In this case soliton are moving along free direction,
transverse to the nonlinearity modulation direction. We consider two
collision scenarios. One occurs when colliding entities are moving
in single rectangular box, and the other when they move along two
different boxes in a double-box modulation profile. In both case we
observe the same diagram of the collision results. For velocities
above a critical value the solitons come out from the collision
being excited - the collision is inelastic. The process of
excitation is accompanied by some radiation. The amplitude of the
excitation grows as we approach critical velocity. If we go below
the critical velocity the solitons do survive the collision, but
soon after they pass though each other, excitation turns into
instability and the soliton gradually dissolve. If we decrease the
velocity even further we eventually enter the region when the
interaction time is long enough for the catastrophic self focusing to arise.

\begin{figure}[tbp]
\includegraphics[width=7cm]{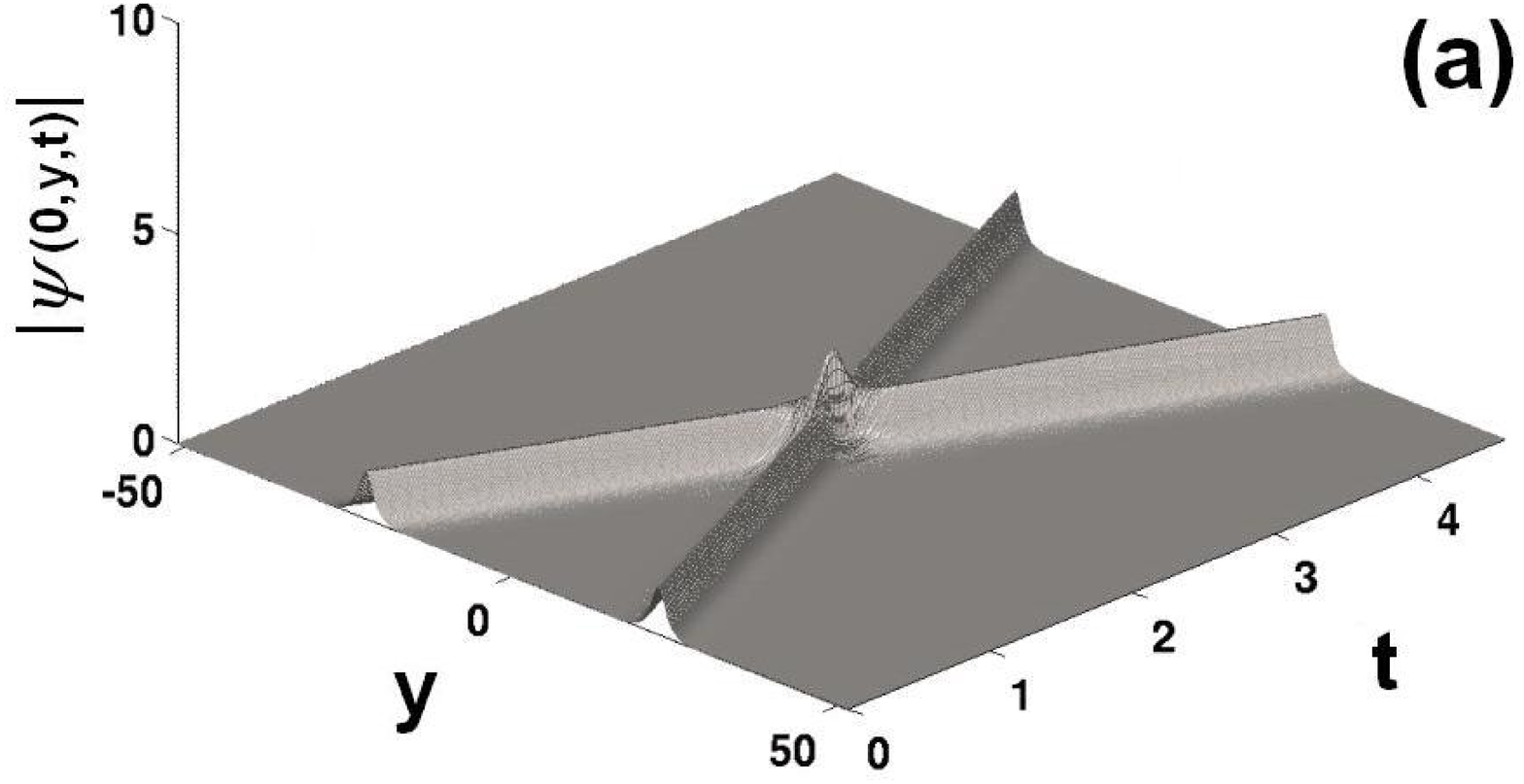}
\includegraphics[width=7cm]{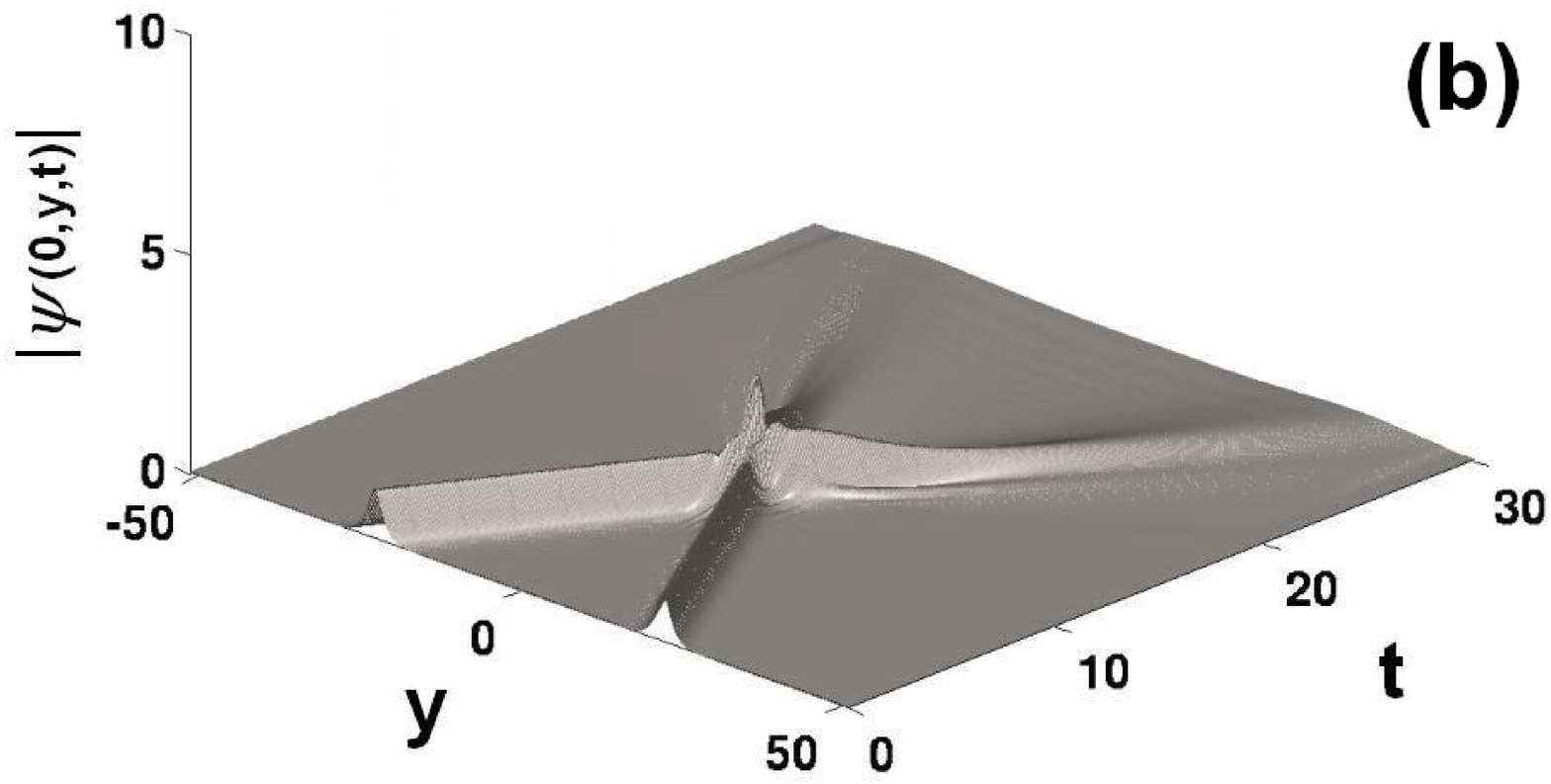}
\includegraphics[width=7cm]{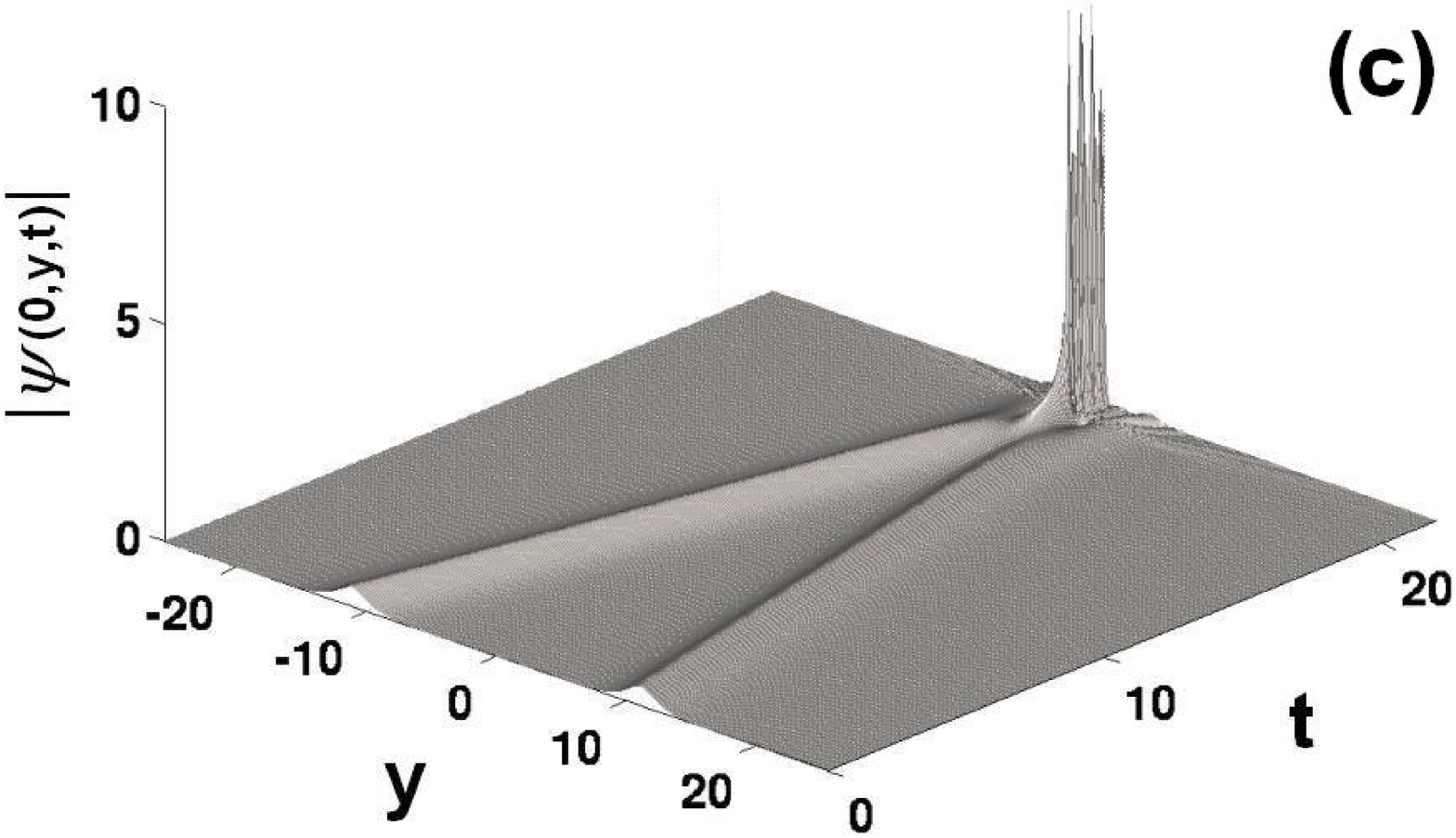}
\caption{Collisions of solitons in  a single box.
Numerical data was obtained for $N=5.8$.  Sections of the wavefunction along $y$ axis are shown for
(a) $v=10$ (excitation), (b) $v=2$ (decay) and (c) $v=0.5$
(catastrophic self focusing) respectively.} \label{galery1}
\end{figure}

\begin{figure}[!htb]
\includegraphics[width=9cm]{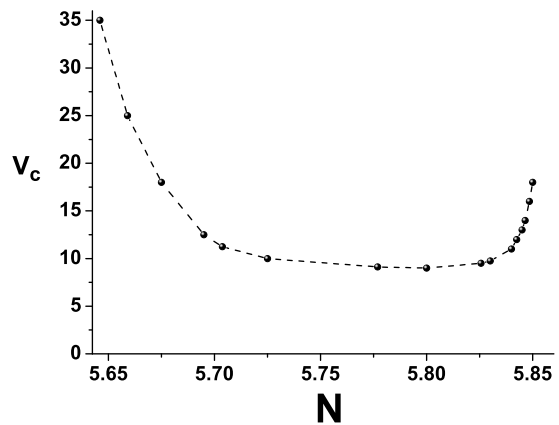}
\caption{Critical velocity versus norm in the case of the single box collision.} \label{hinhgh2}
\end{figure}

\begin{figure}[tbp]
\includegraphics[width=8.5cm]{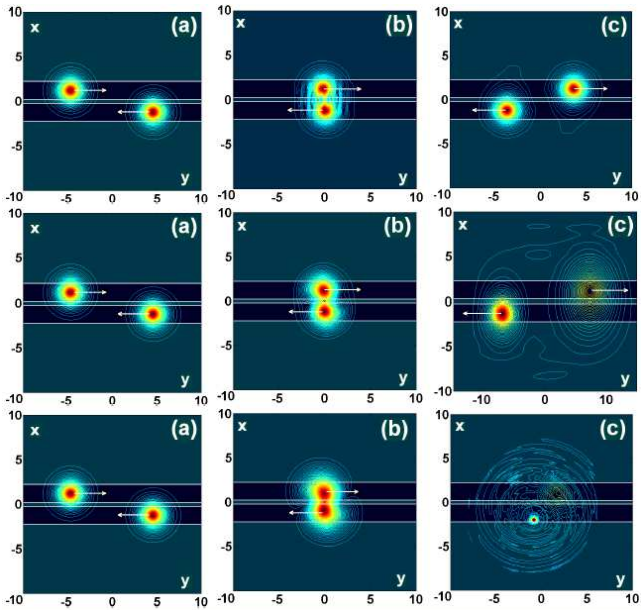}
\caption{Contour plots illustrating soliton collisions in a double box. Numerical results were obtained
for $N=5.8$ and $d=1.2$. Three figures in the each row show contours of
the wavefunction (a) before, (b) during and (c) after the collision.
The top row corresponds to $v=4$. The solitons get excited because of the collision.
The middle row corresponds to $v=1$, where the solitons decay after the collision. Bottom row
corresponds to $v=0.5$, where we observe catastrophic self focusing.} \label{galery2}
\end{figure}

First we considered the case when nonlinearity modulation function is of the form of a single rectangular box (see Eq.~(\ref{one})). Stable
solitons in this case exist only for the value of the norm between
$5.645$ and $5.85$ as we know from previous considerations (see Sec.
III, Fig.~\ref{hin1h}). Three generic type of dynamics described
above are shown in Fig.~\ref{galery1}. The plot of critical
velocity versus norm of each of the soliton is presented in Fig.~\ref{hinhgh2}. The nonmonotonic behavior can be understood as know
that single soliton is least stable at the borders of its existence
region. In the case of double-box modulation profile we in principle
see the same kind of phenomena. We collide two asymmetric solitons
which principal part is occupying different channels. The results are
presented in Fig.~\ref{galery2}.

\section{Conclusions}

In this work, we have considered 2D solitons in the model of the medium with
the self-attractive nonlinearity whose strength is subject to the spatial
modulation in the form of a single or double stripe. The model may be
implemented in BEC and nonlinear optics. It is known that the stabilization
of 2D solitons by means of the spatial modulation of the self-attractive
nonlinearity is a challenging problem. Using the combination of the
variational approximation and direct simulations, we have found that the
ability of the stripe-shaped modulation to stabilize the solitons crucially
depends on its shape: while a smooth modulation profile in the form of a
Gaussian does not support any stable 2D soliton, the rectangular
(box-shaped) profile, as well as its double-box counterpart, give rise to
stable solitons. In the case of the double stripe, three species of stable
soliton solutions were found, \textit{viz}., the single-peak symmetric and
asymmetric solutions, and double-peak symmetric ones. Single-peak solitons
of both symmetric and asymmetric types are accurately described by the VA,
and their stability is adequately accounted for by the VK criterion. The
symmetric double-peak solitons were found in a numerical form, turning out
to be stable in a small region. Collisions between stable solitons were
studied too, by means of direct simulations. We studied collisions in single and double rectangular boxes. Depending on the colliding partners velocity we observed three types of behavior: emerging solitons get some excitations for large velocity, for smaller velocity they decay (excitations were large enough to cause instability), and for even smaller velocity we enter the region where catastrophic self focusing occurs.

A challenging direction for the extension of the analysis reported in this
work is to search for stable solitons supported by spatial modulations of
the nonlinearity in the 3D space.

\section{Acknowledgement}

The authors acknowledge support of the Polish Government Research
Grants for years 2007 – 2009 (N. V. H. and M.T.) and for years 2007 – 2010 (P.Z). The work of B. A. M. was supported, in a part, by the German - Israel Foundation through grant No. 149/2006.

\end{document}